\documentclass{nature}
\usepackage{graphicx}
\usepackage{epstopdf,epsf,epsfig,ulem,verbatim}
\usepackage{floatrow,color,soul,lineno}
\title{Characterizing Si:P quantum dot qubits with spin resonance techniques}

\author{Yu Wang$^{1,*}$, Chin-Yi Chen$^{1}$, Gerhard Klimeck$^1$, Michelle Y. Simmons$^2$, \& Rajib Rahman$^1$ }

\begin{document}
\maketitle
\begin{affiliations}
 \item Network for Computational Nanotechnology, Purdue University, West Lafayette, IN 47907, USA
 \item Centre for Quantum Computation and Communication Technology, School of Physics, University of New South Wales, Sydney, NSW 2052, Australia 
\end{affiliations}
\nolinenumbers
\noindent*Correspondence to wang1613@purdue.edu
\newline

\begin{abstract}
Quantum dots patterned by atomically precise placement of phosphorus donors in single crystal silicon have long spin lifetimes, advantages in addressability, large exchange tunability, and are readily available few-electron systems. To be utilized as quantum bits, it is important to non-invasively characterise these donor quantum dots post fabrication and extract the number of bound electron and nuclear spins as well as their locations. Here, we propose a metrology technique based on electron spin resonance (ESR) measurements with the on-chip circuitry already needed for qubit manipulation to obtain atomic scale information about donor quantum dots and their spin configurations. Using atomistic tight-binding technique and Hartree self-consistent field approximation, we show that the ESR transition frequencies are directly related to the number of donors, electrons, and their locations through the electron-nuclear hyperfine interaction.

\end{abstract}

\flushbottom

%
%
\thispagestyle{empty}

\section*{Introduction}

Silicon is an excellent host for spin qubits as it is a spin-free environment with the added advantage of weak spin-orbit coupling. The coherence time of single electron spins bound to phosphorus donors in enriched silicon-28 was measured to be 0.6 s with Hahn echo in bulk\cite{Tyryshkin}. Recent advances in scanning tunneling microscope (STM) based lithography\cite{Schofield} have enabled the realization of donor-based QDs buried deep inside silicon far from hetero-interfaces by atomically precise placement of phosphorus donors\cite{Martin1}. Electrons bound to these Si:P QDs are confined not by a surface gate potential that depletes the surrounding two dimensional (2D) electron gas, but by the 3D Coulombic potential of a few donors located within $\sim$nanometers of each other, and form readily available few-electron systems. These donors can be addressed by in-plane gates also fabricated by STM lithography\cite{Buch}. Already, transport spectroscopy has been performed on single, double, and triple QDs of Si:P\cite{Martin2, Weber, Watson}, and single-electron spin readout and two-electron spin blockade have been demonstrated\cite{Buch, Weber}. Theoretically, it has been shown that Si:P quantum dots can have exceptionally long spin relaxation times\cite{Hsueh}, and advantages in addressability\cite{Buch} as well as large exchange tunability in multi-qubit devices\cite{YWang}. A surface code based potentially scalable architecture has also been proposed recently with STM patterned Si:P dots\cite{Hill}. Spin coherence measurements based on donor ensembles also provide some evidence that buried donors may have better coherence times than donors close to interfaces \cite{Schenkel}, which is likely due to deleterious noise sources present near the interface\cite{Paik}.    

The popular Kane qubit proposal of single phosphorus based quantum computer\cite{Kane} utilizes an ac-magnetic field for single qubit operations and an inter-donor exchange coupling for two-qubit operations. Experimentally, ac-magnetic fields have been used to perform electron spin resonance (ESR) and nuclear magnetic resonance (NMR) on single phosphorus electron\cite{Pla1} and nuclear spins\cite{Pla2} respectively. Recently, a two-qubit logic gate in gate-defined silicon QDs has also been demonstrated with ESR based single qubit control\cite{Menno2}. To utilize Si:P QDs as spin qubits, 
it is important to know the precise number of electron and nuclear spins bound to these dots, which can also help to identify appropriate spin transitions of the qubit and to deplete the dots down to the ideal regime of one electron occupation. Understanding the extent of the electron wavefunctions of the donor dots is important for optimizing the inter-dot exchange and tunnel coupling useful for two-qubit gates. In this work, we present a metrological method based on a non-invasive atomic-scale characterization technique to extract information about the donor numbers, electron numbers, and donor locations based on ESR measurements using the on-chip circuitry already needed for qubit manipulation. The method relies on probing the spin splittings of the donor dots through the electron-nucleus hyperfine coupling which is sensitive to atomic-scale details of the donor dot. To establish this metrology technique, we compute hyperfine couplings as a function of donor number, electron number, and donor locations using an atomistic tight-binding method with a Hartree approximation, and show how the combined experimental characteristic with theoretical mapping can inform the use of donor dot qubits.

Although transport spectroscopy can also be used to determine the number of electrons bound to a donor dot\cite{Weber} based on charging energy extraction for various charge transitions, such a procedure has some associated uncertainties as the charging energies depend on the electrostatics and geometry of the device and can show large variations between devices\cite{Weber}. Moreover, electron transport depends on a complicated interplay of a number of parameters such as lead-to-dot tunnel couplings, dot energy levels, charging energies, spin and charge relaxation rates\cite{Martin1, Martin2}. As a result, it is often difficult to ensure that the last charge transition measured in the device indeed corresponds to a single electron occupation. Also, there are uncertainties in donor locations up to 1 nm resulting from donor incorporation mechanism, donor diffusion and segregation within the dot and in the leads\cite{Joris}. All these effects produce a wide band of charging energies as shown in Ref\cite{Weber}, making the extraction of electron and donor numbers difficult. The proposed ESR based metrology circumvents these challenges and provides a more accurate way to characterize and customize Si:P QDs for spin qubit experiments.

\begin{figure}[htbp]
\center
\epsfxsize=5in\epsfbox{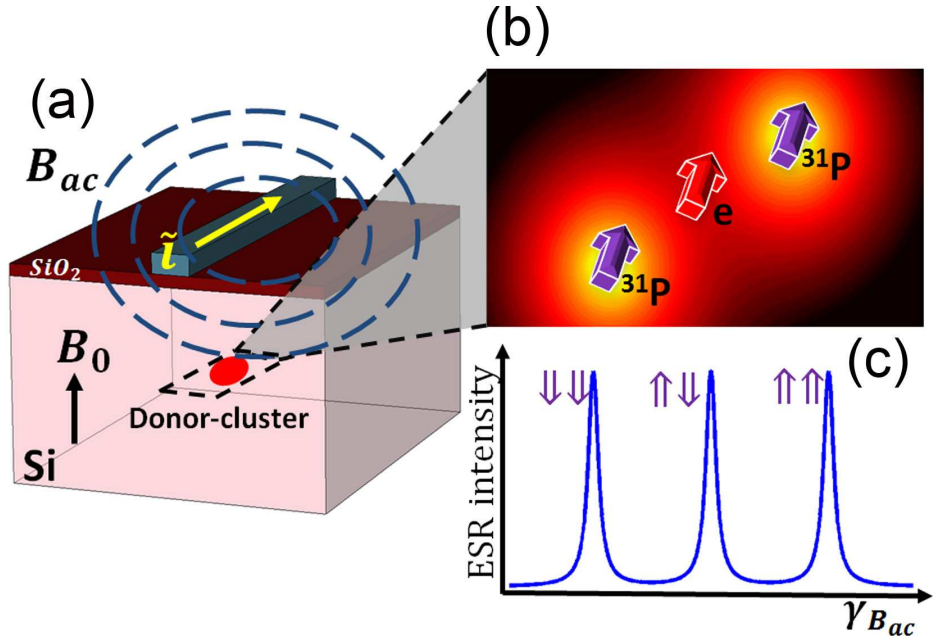}
\caption{(a) A schematic of a donor dot qubit device in silicon with ESR measurements using an ac-magnetic field ($B_{ac}$). (b) The potential landscape of a 2P donor-dot with one bound electron (2P:1e), where light color represents high magnitude and dark color represents low magnitude. The arrows indicate the nuclear and electronic spins. (c) The schematic of a representative ESR measurement on 2P:1e showing three transition peaks associated with given nuclear spin configurations (thick arrows).}
\vspace{0cm}
\label{fi1}
\end{figure}

First, we will discuss how this metrology works theoretically on different device samples. Fig. 1(a) shows a simplified schematic of a donor dot qubit with a microwave ESR line. In reality, this metrology is most feasible for donor-qubit devices with ESR-addressability\cite{Pla1, Kane_A_Gate}. The donor dot (shown as the red disk in the dashed rectangular region) is patterned into the silicon substrate (the pink box) by STM lithography\cite{Buch}. A static magnetic field ($B_0$) is applied to the system to polarize the electron spins bound to the dot. The metal wire (the blue bar) deposited on the SiO$_2$ layer carries an alternating current that can generate an electromagnetic field ($B_{ac}$ represented by the blue dashed ellipses) to rotate the qubit electron spin with microwave frequency. Spin readout can be performed by spin dependent tunneling to a single-electron-transistor using the protocol described in Refs\cite{Buch, Morello}. 

Fig. 1(b) shows an electron spin (red arrow) bound by the quantum confinement of a 2P donor dot containing two nuclear spins (purple arrows). The background depicts its electrostatic potential landscape. The spins are coupled by the electron-nucleus hyperfine interaction which depends on both the nuclear spin orientation and the electronic probability density at the site of the nuclear spins. The latter is sensitive to the quantum confinement of the electron wavefunction provided by the potential of the donors, and hence, in the more general case, depends on the number of donors and electrons, as well as their locations. This dependency is crucial for the metrology technique proposed here as shown later. A solution of the electron-nucleus spin Hamiltonian with hyperfine interaction, magnetic dipole interaction, and an external B-field provides specific up/down transition frequencies at which we can address the electron spin. 

Fig. 1(c) shows a schematic of three ESR transition frequencies that could exist for a 2P:1e dot associated with different possible orientations of the two nuclear spins (thick purple arrows). As the overall hyperfine interaction depends on the nuclear spin orientations, three possible ESR frequencies can be detected due to three possible nuclear spin configurations in a 2P donor dot. The splittings between the ESR peaks depend on the strength of the hyperfine interaction. Experimentally the transition frequencies can be resolved easily since the broadening of the peaks in the ESR spectrum are $\sim$8 MHz in natural silicon{\cite{Pla1}}, and can be reduced to $\sim$1.8 kHz in enriched silicon-28 substrate\cite{Muhonen}. The separations between the peaks are in the order of 100 MHz for most Si:P QDs as determined by the hyperfine interaction{\cite{Pla2}}. An experiment-theory comparison helps to map the number of peaks and their locations in the ESR spectra to specific configurations of the Si:P QD.

\section*{Methods}

To accurately calculate the ESR frequencies of a Si:P QD, it is crucial to capture its electronic structure. Here we obtain the ESR frequencies by solving the effective spin Hamiltonian based on the atomistic tight-binding approach. A donor-dot consists of $m$ nuclear spins and $n$ electron spins localized in a silicon crystal lattice. In spin qubit experiments, a static external magnetic field $\vec{B}$ is applied to generate well-defined spin polarized states split by the Zeeman energy. Its effective spin Hamiltonian thereby can be expressed as
\begin{equation}  
H_{spin} = H_{zeeman}+H_{hyperfine}. \label{eq1}
\end{equation}
The external B-field is included in the Zeeman term, $H_{Zeeman}$, as  
\begin{equation}  
H_{zeeman}=\sum_{i=1}^{n} g_e \mu_B \vec{B} \cdot \vec{S_i}-\sum_{j=1}^{m} g_n \mu_n \vec{B} \cdot \vec{I_j}, \label{eq2}
\end{equation}
where $\vec{S_i}$ and $\vec{I_j}$ denote the spin operator of the $i$th electron and the $j$th nucleus respectively. $g_e$ and $g_n$ are the electron and nuclear g-factors respectively. $\mu_B$ is the Bohr magneton and $\mu_n$ is the nuclear magneton. The second term, $H_{hyperfine}$, is the hyperfine interaction between electron spins bound to the donor-dot and the nuclear spins of the donors, and is given as, 
\begin{equation}  
H_{hyperfine}=\sum_{i}^{n} \sum_{j}^{m} A_{ij} \vec{I_j} \cdot \vec{S_i}+\sum_{i}^{n} \sum_{j}^{m} \vec{I_j} \cdot \widetilde{\widetilde{D_{ij}} }\cdot \vec{S_i}, \label{eq3}
\end{equation}
where $A_{ij}$ represents the Fermi contact hyperfine coupling of the $i$th electron and the $j$th nuclei, which is proportional to the electronic probability density at the $j$th donor site. $\widetilde{\widetilde{D_{ij}}}$ is the anisotropic hyperfine (or magnetic dipolar) interaction tensor. $A_{ij}$ and the components of $\widetilde{\widetilde{D_{ij}}}$ can be expressed as \cite{Hale_Mieher, Park}: 
\begin{equation}
A_{ij} = \frac{\mu_0}{4\pi}g_e\mu_Bg_n\mu_n\frac{8\pi}{3}|\Psi_i(\vec{R_j})|^2, \label{eq4}
\end{equation}
\begin{equation}
D_{ij,kl} = \frac{\mu_0}{4\pi}g_e\mu_Bg_n\mu_n\langle\Psi_i|\frac{3r_kr_l-|\vec{r}-\vec{R_j}|^2\delta_{kl}}{|\vec{r}-\vec{R_j}|^5}|\Psi_i\rangle, \label{eq5}
\end{equation}
where $\mu_0$ is the vacuum permeability. $\Psi_i$ is the wavefunction of the $i$th electron, $\vec{R_j}$ is the position vector of the $j$th nucleus, and $r_{k,l} = (x,y,z)$ are the electron position vector components. Although we consider both contact and the anisotropic hyperfine interactions in the simulations, we find that the $\widetilde{\widetilde{D_{ij}}}$ terms are at least three orders of magnitude less than the $A_{ij}$ terms, analogous to the case of Si$^{29}$ in silicon\cite{Hale_Mieher, Park}. Hence, the spin splitting is mostly affected by the $A_{ij}$ terms. This helps in the direct mapping of ESR frequencies to $A_{ij}$ and eventually to the physical properties of the donor dot, as needed for the metrology. In obtaining the $A_{ij}$ coupling from the tight-binding wavefunction, we adopt the same technique as Ref\cite{Rahman} using the experimentally measured hyperfine resonance frequency of 117.53 MHz for a bulk phosphorus as a calibration.

To obtain the electron wavefunctions, the tight-binding Hamiltonian of the silicon and the donor atoms is represented by an sp$^3$d$^5$s* spin-resolved atomic orbital basis with nearest neighbor interaction and spin-orbital coupling. The resulting eigenvalue problem is solved by a parallel block Lanczos algorithm for about 1.4 million atoms using the nanoelectronic modeling tool (NEMO-3D)\cite{Klimeck}. The Hamiltonian contains three terms as shown below.
\begin{equation}
H = H_{Si} + H_D+H_{e-e}, \label{eq7}
\end{equation}
where $H_{Si}$ represents the Hamiltonian of the silicon lattice, $H_D$ denotes the central-cell corrected Coulomb potential energy given by the positive charges of donor cores, and $H_{e-e}$ is the mean-field electron-electron interaction energy present for multi-electron occupation of the donor dot. The term $H_{Si}$ contains the full bandstructure of silicon\cite{Boykin} along with hydrogen passivated surface atoms\cite{Lee}. Each donor potential in $H_D$ is represented by a Coulomb potential screened by the dielectric constant of silicon along with a central-cell correction term which assumes a constant potential $U_0$ at the donor site\cite{Rahman}. This model has been well calibrated with single donor spectroscopy measurements and reproduces all the bound states of the donor with correct energy levels\cite{Shaikh}. The donor wavefunction obtained from this model also agrees very well with measurements from recent STM imaging experiments\cite{Salfi}. For multi-electron occupation, a mean-field method is utilized to reduce the computational complexity using the Hartree self-consistent field (SCF) approximation. In this scheme, the mean-field potential energy of the electrons is solved self-consistently with the Poisson equation, assuming that the electrons occupy the lowest states conforming to the Pauli exclusion principle. This method has successfully captured the experimental two-electron binding energy of a single donor in silicon called the D$^-$ state\cite{Rahman2}, multi-electron binding energies in donor dots\cite{Weber}, and can also reproduce experimentally measured spin-lattice relaxation times $T_1$ times in donor dots\cite{Hsueh}, providing us confidence in the mean-field approximation for multi-electron occupation.

As for multi-electron occupation, for qubit applications, we are mostly interested in effective 1/2 spin configurations of a donor dot, where there is an odd numbered electron in the dot. As indicated by eq. (\ref{eq3}) and (\ref{eq4}), the net hyperfine interaction at site $\vec{R_j}$ depends on the on-site spin density  $\sum_{p=\uparrow}\vert\Psi_p(\vec{R_j})\vert^2-\sum_{q=\downarrow}\vert\Psi_q(\vec{R_j})\vert^2$. The inner electrons form spin pairs occupying the same orbital states of the dot. As a result, they have identical spatial wavefunction but different spins $S_i$ which leads to no contribution to the hyperfine term  $\sum_{i}^{n} \sum_{j}^{m}A_{ij}\vec{I_j}\cdot\vec{S_i}$. Thereby, the hyperfine interaction as well as the ESR frequencies of donor-dot with multi-electron occupation only depends on the wavefunction of the unpaired outermost electron, which can be captured by the mean-field theory applied here.

\section*{Results and Discussions}

In the results presented in the following, the simulation domain is set to 30nm $\times$ 30nm $\times$ 30nm of a regular silicon crystal lattice ($\sim$1.4 million atoms) in NEMO-3D, so that the Coulombic potential of the donors approach almost zero near the boundaries of the domain. This ensures that the donor-dot bound wavefunctions are not affected by the artificial hard-wall confinement of the box of silicon. The P-donor dots are placed in the middle of the box in the (001) atomic plane. A static magnetic field $B_0$ = 1.5T is applied in the [001] direction.

\begin{figure}[htbp]
\center\epsfxsize=5in\epsfbox{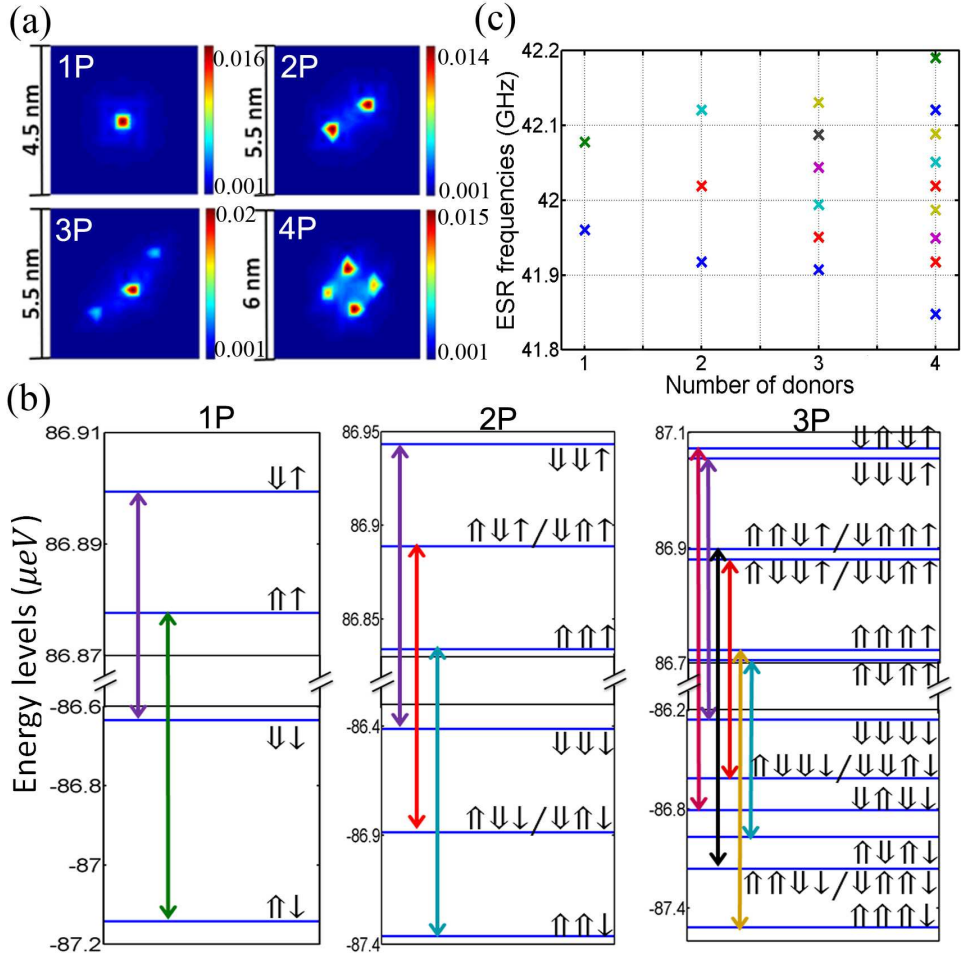}
\caption{ Dependence of ESR spectra on donor number in donor-based quantum dots. (a) Single-electron probability density distribution of the ground states (in nm$^{-3}$). The four figures show the 1-4 P donor cases respectively. (b) The electron spin transitions of devices with 1-3 P donors corresponding to each spin configuration of electron $\uparrow$ or $\downarrow$ and nuclei $\Uparrow$ or $\Downarrow$. The thick (thin) arrows indicate the orientations of the nuclear (electron) spins. The ground orbital state energy without B-fields is the zero energy reference. (The 4P case follows the same routine and is not shown here as 2$^5$ spin configurations are involved.) (c) The ESR frequencies of devices with 1-4 P donors.}
\vspace{0cm}
\label{fi2}
\end{figure}

First, we show how we can infer the number of donors in a donor-dot using ESR measurements. Fig. 2(a) shows the single-electron probability density distributions for 1 to 4 P donors in the central plane. Fig. 2(b) shows the energy levels computed from the spin Hamiltonian (eq. (\ref{eq1})) for single electron occupation. The levels are shifted relative to the ground orbital state located at zero energy for reference due to Zeeman effect. The number of energy levels increases with the number of donors in a dot, as the total number of nuclear spins increases. The spin orientations of the nuclear and electron spins can be identified from the eigenvectors of the spin Hamiltonian, and are shown in Fig. 2(b) as well. The transition frequencies for an electron spin flip can be computed from the difference between the up and down electron spin (thin arrows) states with the same nuclear spin orientations (thick arrows). Fig. 2(c) shows the ESR transition frequencies for the 4 cases. These transition frequencies correspond to the location of the peaks in a sample ESR measurement as illustrated in Fig. 1(c). Firstly, as expected the number of peaks increases with the number of donors since the number of possible nuclear spin configurations increases. If the ac-B-field frequency can be controlled with enough accuracy, all the ESR peaks can in principle be observed. Comparing the number of measured peaks with the theoretical calculations of Fig. 2(c), it is possible to determine the number of donors in the quantum dot.

Not only are the number of ESR transitions important, but their locations and spacings in the frequency axis can also provide information about the locations of the P donors and the number of bound electrons in the quantum dot. In the next section, we discuss the effect of the donor locations on ESR frequencies. The hyperfine couplings between the electron and the nuclear spins, which are proportional to electron probability density at each donor site, are quite sensitive to the relative donor locations within the quantum dot. The quantum confinement and hence, the on-site wavefunction concentrations strongly depend both on the radial and angular separation of the donors due to the crystal symmetry, even for $\sim$1-2 nm variations in quantum dot sizes. It is critical to obtain information about the extent of the electron wavefunction for the design of two-qubit gates\cite{YWang} and to engineer long $T_1$ times\cite{Hsueh}.  Moreover, it is useful to obtain a range within which hyperfine couplings can vary given the number of donors in a dot. 

\begin{figure}[htbp]
\center\epsfxsize=5in\epsfbox{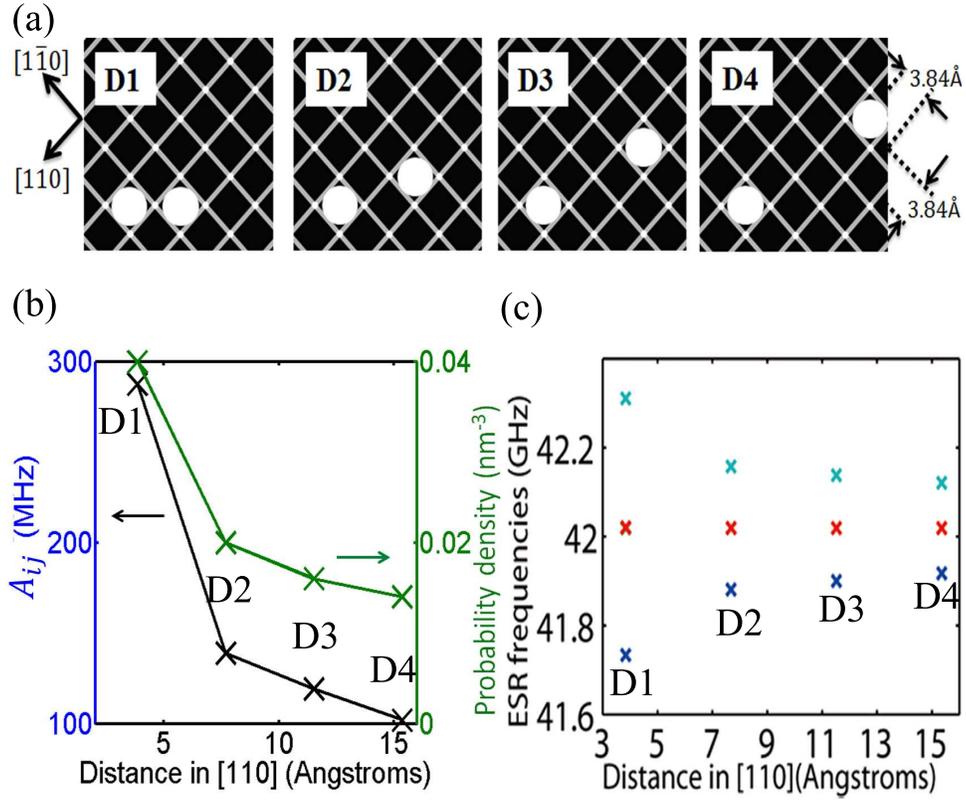}
\caption{ Dependence of ESR spectra on donor separation with the donor-based quantum dot. (a) The four studied positional configurations of 2P donor-dots (D1-D4) respectively. The black squares represent silicon atoms and white filled circles represent the substituting P atoms on an atomic plane of silicon. The nearest-neighbor atoms are separated by $\sim$3.84 $\AA$ in the [110] or [1$\bar{1}$0] direction. (b) The Fermi contact hyperfine constant $A_{ij}$ as a function of two donor separation in [110] corresponding to (a). (c) The impact of variations in the Fermi contact hyperfine couplings on the ESR frequencies of D1-D4 cases in (a). Here we find variations in ESR frequencies up to $\sim$190 MHz.}
\vspace{0cm}
\label{fi3}
\end{figure}

In a STM patterned Si:P quantum dots, a lithographic patch defines a region where donors can be incorporated into the silicon crystal. The donors in reality can occupy different energetically favourable atomic sites within the patch, which provides some variations in hyperfine couplings between dots with the same number of donors. Here, we use 2P donor dots as an example to demonstrate how the hyperfine couplings and hence the ESR frequencies are correlated with donor separations within a 2P donor quantum dot. Fig. 3(a) shows four different 2P configurations within a lithographic patch on the silicon (100) plane, labeled as D1-D4. The black squares represent silicon atoms and the white filled circles represent the substituting P atoms on a (001) atomic plane of silicon crystal lattice. From left to right, the separation of the two donors are increased by a step of $\sim$3.84 angstroms in the [110] direction, from D1 to D4. And in the [1$\overline{1}$0] direction, two donors are $\sim$3.84 angstroms apart, which is fixed for all the four cases. The right y-axis of Fig. 3(b) shows that the on-site electron probability density decreases with the distance between two donors.  This will lead to a decrease in hyperfine coupling with the distance, which is shown as the left y-axis in Fig. 3(b). Fig. 3(b) gives the values of Fermi contact hyperfine coupling ($A_{ij}$) of one of the two donors, in which the other shares the same value because of symmetry. Fig. 3(c) shows the corresponding ESR frequencies of these four configurations. We note the ESR frequencies are farther apart when the donors are closer spatially. The results provide valuable information about the extent of the electron wavefunctions in devices where we have the same number of donors but different donor locations.

\begin{figure}[htbp]
\center\epsfxsize=5in\epsfbox{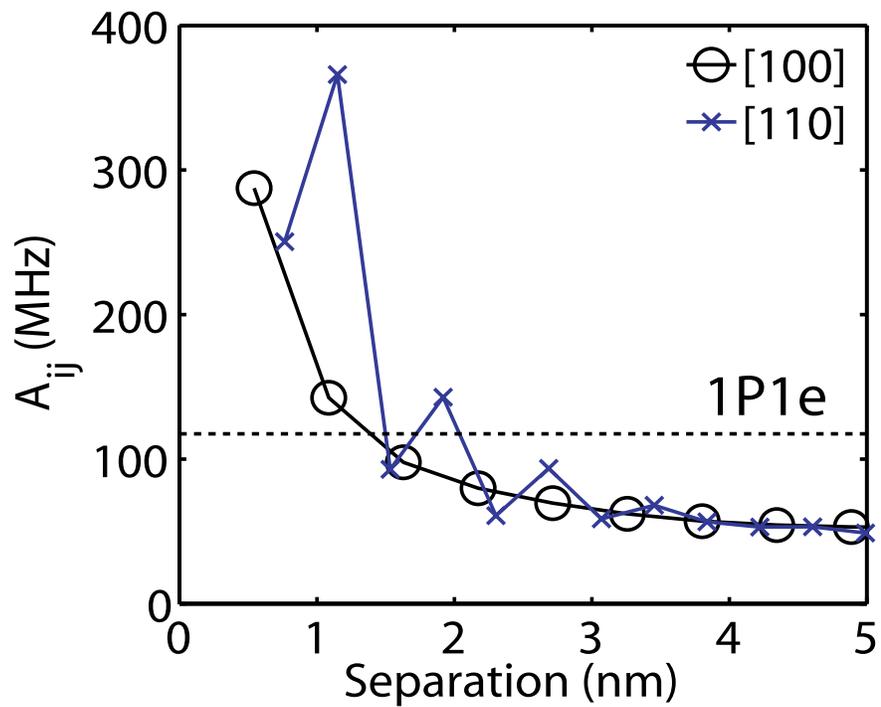}
\caption{ Impact of inter-donor separation along the [100] and the [110] direction on the Fermi contact hyperfine coupling $A_{ij}$ of a donor dot. The horizontal dashed line represents the hyperfine coupling of a single P donor in bulk silicon, which is $\sim$117.53 MHz.}
\vspace{0cm}
\label{fi4}
\end{figure}

Effects such as dopant diffusion and segregation can often cause the precisely placed donors to occupy lattice sites outside the lithographic patch. To account for this scenario, we explore the hyperfine interaction of a P donor pair for larger radial separations along two high symmetry directions in the plane. Fig. 4 shows the Fermi contact hyperfine coupling $A_{ij}$ as a function of inter-donor separation along the [100] (circles) and the [110] direction (crosses). In the [100] direction, $A_{ij}$ first decreases from 287.4 MHz to 57.1 MHz from a separation of $\sim$0.54 nm to $\sim$3.8 nm, and smoothly levels off when the separation is larger than 4 nm. A very close separation distance increases the quantum confinement of the electron wavefunction as the potential of the two donors enhance each other significantly near the donor cores. Here the electron density from the outskirts of the dot is redistributed to the central region forming a strongly hybridized molecular state, thereby enhancing the hyperfine coupling.

When the separation is even larger ($>\sim$5 nm), the two P donors approach the regime of two isolated bulk-like donors with $A_{ij}$ converging to about half of the bulk value of 117.53 MHz, as the two donors still share one electron. This is in contrast to the case of two far separated donors with one electron each where $A_{ij}$=117.53 MHz. In the [110] direction, $A_{ij}$ can be seen to oscillate as a function of separation due to the interference between the wavefunctions of the six conduction band valleys of silicon, similar to the predicted exchange oscillations\cite{Koiller}. From Fig. \ref{fi4} we can see that $A_{ij}$ of a P-donor pair in silicon with one bound electron can vary from $\sim$366.0 MHz to $\sim$48.9 MHz within a 5 nm separation range.

Next, we evaluate the ESR frequencies of the donor-based quantum dots for multi-electron occupation. As discussed earlier, the resultant hyperfine coupling in a donor dot originates from the unpaired electronic spin of the outermost electron. In such a case, the unpaired electron is screened from the positively charged donor cores by the spin-paired electrons of the inner orbitals, and the wavefunction spreads out more in space causing a reduction in the density at the nuclear sites in the central region. This causes a sharp decrease in the Fermi contact hyperfine coupling as more electrons are loaded. Note that, by referring to spin-paired inner electrons, we are still referring to the extra electrons in the donor molecules that do not take part in the covalent bonds with the nearest silicon atoms as these bonds remain intact in the donor-dot.

\begin{figure}[htbp]
\center\epsfxsize=5in\epsfbox{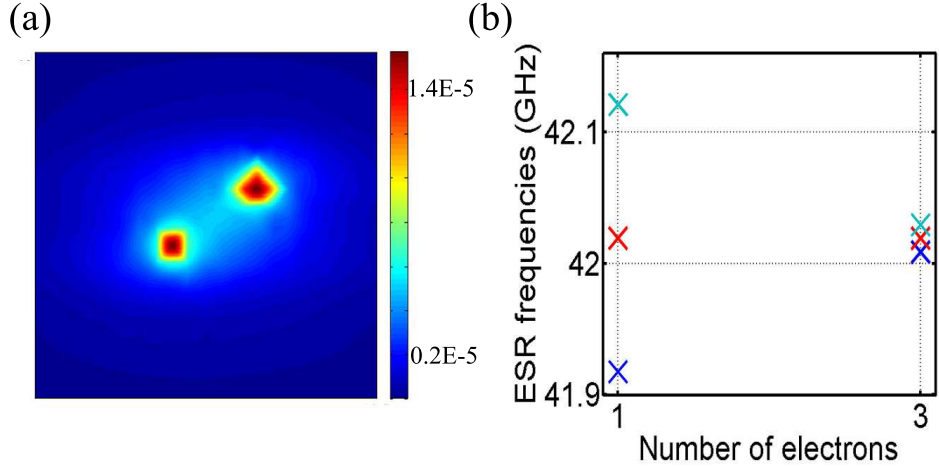}
\caption{ Impact of multi-electron occupation in a donor-dot on ESR  frequencies. (a) Single-electron probability density distribution of the 3$^{rd}$ electron in D4 with Hartree self-consistent field solution (in nm$^{-3}$). (b) The ESR frequencies of the 3$^{rd}$ electron in D4, compared with the case of only 1 electron in D4 (shown as the 2$^{nd}$ figure in Fig. 2(a)).}
\vspace{0cm}
\label{fi5}
\end{figure}

In Fig. 5(b), we show the ESR frequencies for 1 and 3 electrons on a 2P donor dot, assuming the D4 positional configuration of Fig. 3(a) in which the donors are separated by $\sim$1.5 nm in the [110] direction. Fig. 5(a) shows the probability density distribution of the 3rd electron solved with Hartree self-consistent field approximation as illustrated in the methodology section. It has a larger extent and lower on-site concentrations compared with the single-electron probability density of D4 (shown as the 2$^{nd}$ figure in Fig. 2(a)). With two more electrons, consequently the Fermi contact hyperfine coupling decreases to 10.5 MHz from 101.8 MHz. Correspondingly, the ESR resonance peaks of Fig. 5(b) are closer to each other if the number of bound electrons in a given donor-dot is larger. Interestingly we observe the effect of multi-electron occupation in hyperfine coupling reduction (a factor of $\sim$10) is more prominent than increasing the inter-donor distance up to $\sim$1.5 nm within the dot (a factor of $\sim$3, Fig. \ref{fi3}(b)). Therefore, the electron number in a donor quantum dot can also be detected with this metrology. To summarize the metrology, the number of ESR frequencies indicates the number of donors within a quantum dot, and the separations between the ESR frequencies imply the bound electron number and inter-donor distances. 

\section*{Conclusion}

We have proposed a metrology technique to characterize spin qubits hosted by Si:P quantum dots in silicon. The metrology can help extract information about the number of donors, donor locations and the number of electrons based on ESR measurements, and is potentially more accurate than charging energy based metrology. An atomistic tight-binding method is coupled with a Hartree self-consistent field approach to obtain the orbital wavefunctions of multi-donor quantum dots in silicon. Solving a multi-spin Hamiltonian with parameters computed from the wavefunctions enables us to predict ESR transition frequencies for the Si:P quantum dots. We found that the number of ESR frequencies increases with the number of donors, while the locations and the spacings of the ESR peaks on the frequency axis can help extract information about the spatial extent of the donors within the quantum dot and the number of bound electrons. The extracted information is useful for engineering single and multi-qubit operations in a system with very long coherence times. The method proposed here may also be applied to obtain information about other $\frac{1}{2}$-spin qubit systems like phosphorus donors in germanium, or non-$\frac{1}{2}$-spin qubit systems, for example, arsenic/boron dopants in silicon, and potentially for dopants in other semiconductors.  

\section*{Acknowledgments}

This research was conducted by the US National Security Agency and the US Army Research Office under contract No. W911NF-08-1-0527. Computational resources on nanoHUB.org, funded by the NSF grant EEC-0228390, were used. 

\section*{Author contributions statement}

Y.W. and R.R. conceived the metrology. Y.W. and C.C. conducted the simulations. Y.W., M.Y.S and R.R. analyzed the results. All authors reviewed the manuscript.

\section*{Additional information}

The authors declare that they have no competing financial interests.

\end{document}